# *Exploring Quantum Active Learning for Materials Design and Discovery*

Maicon Pierre Lourenço[1*], Hadi Zadeh-Haghighi[2], Jiří Hostaš[3,4], Mosayeb Naseri[4], Daya Gaur[5], Christoph Simon[2] and Dennis R. Salahub[4]

1- Departamento de Química e Física – Centro de Ciências Exatas, Naturais e da Saúde – CCENS – Universidade Federal do Espírito Santo, 29500-000, Alegre, Espírito Santo, Brasil.
2- Department of Physics and Astronomy, Institute for Quantum Science and Technology, Quantum Alberta, and Hotchkiss Brain Institute, University of Calgary, Calgary, AB T2N 1N4, Canada.
3- Digital Technologies Research Centre, National Research Council of Canada, 1200 Montréal Road, Ottawa, ON, K1A 0R6 Canada.
4- Department of Chemistry, Department of Physics and Astronomy, CMS Centre for Molecular Simulation, IQST Institute for Quantum Science and Technology, Quantum Alberta, QHA Quantum Horizons Alberta, University of Calgary, 2500 University Drive NW, Calgary, AB, T2N 1N4, Canada.
5- Department of Computer Science, University of Lethbridge, 4401 University Dr. West Lethbridge, AB T1K 3M4, Canada.

*Address correspondence to:* maiconpl01@gmail.com (MPL).

# Table of Contents

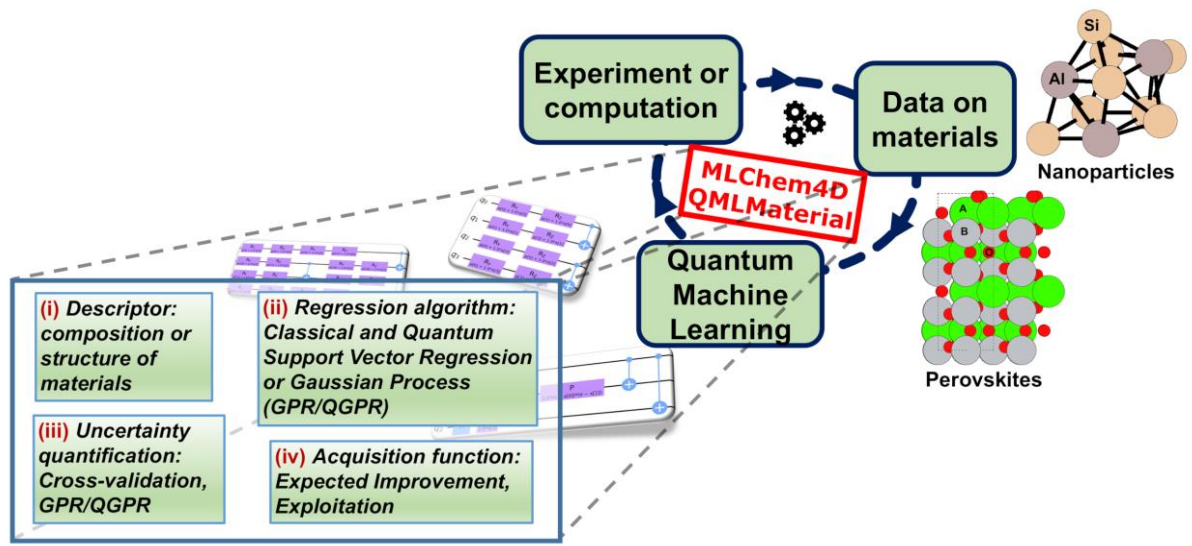

## *Abstract*

The meeting of artificial intelligence (AI) and quantum computing is already a reality; quantum machine learning (QML) promises the design of better regression models. In this work, we extend our previous studies of materials discovery using classical active learning (AL), which showed remarkable economy of data, to explore the use of quantum algorithms within the AL framework (QAL) as implemented in the MLChem4D and QMLMaterials codes. The proposed QAL uses quantum support vector regressor (QSVR) or a quantum Gaussian process regressor (QGPR) with various quantum kernels and different feature maps. Data sets include perovskite properties (piezoelectric coefficient, band gap, energy storage) and the structure optimization of a doped nanoparticle ($3Al@Si_{11}$) chosen to compare with classical AL results. Our results revealed that the QAL method improved the searches in most cases, but not all, seemingly correlated with the "roughness" of the data. QAL has the potential of finding optimum solutions, within chemical space, in materials science and elsewhere in chemistry.

## *Introduction*

Quantum computing (QC) and quantum machine learning (QML) present exciting prospective applications of quantum technologies[1,2], in the Noisy Intermediate-Scale Quantum[3] (NISQ) technology era and beyond, with flourishing new discoveries and perspectives[4]. In QML applications, quantum computers can aid in speeding up machine learning (ML) training[2] and providing quantum enhancement of ML inference, allowing better generalization[5]. Thus, as the overlap of AI and quantum computing is already a reality, we might be closer to the *Quantum Artificial Intelligence* (QAI) era, where QML models can aid in the development of better agents[6] through quantum kernel[2] or quantum neural networks[7] methods.

Being able to predict the next experimental conditions to find new materials with improved properties – such as piezoelectric coefficients[8,9], band-gaps[10,11] and energy storage densities[8,12] – is challenging since this requires searching through uncharted chemical spaces[13], mostly by taking into account only limited data. Data acquisition is time consuming, expensive and error prone: whether the data is obtained from synthesis and characterization in the laboratory or from computational modeling and simulation. This encourages the development of smart agents[14] to help in decision tasks. With complex multi-element oxides, such as the perovskites that are widely used in energy applications[15], the intuition gained by studying related binary oxides is not always applicable due to the emergence of unexpected properties and behavior[16-19], and an alternative method is needed to provide the necessary guidance regarding promising compositions for the specific end-application. Therefore, dealing with small datasets is a reality in chemistry and materials science and methods that assist with optimal experimental design, guided by data, are crucial.

In this work, we extend our previous work using classical active learning (AL) and propose a quantum active learning (QAL) approach – to guide new experiments[20] or computations for optimum experimental design or automatic structural determination[21], with as little data as possible. QAL is based on quantum supervised machine learning (ML), in particular quantum support vector regression[22] (QSVR) and quantum Gaussian process regressor[23] (GPR), where the epistemic uncertainty is obtained from K-fold cross-validation[24] (CV) and analytically using Bayesian statistics[25,26], respectively. This uncertainty is used to evaluate acquisition or utility functions for decision-making[27], i.e., to choose the next experimental conditions for further experiments from the unobserved search space. More data is acquired during the QAL cycle and the QML regression is improved, increasing the probability of finding the optimum properties of the materials or experimental conditions with as few as possible new laboratory or computational experiments.

This study presents the QAL approach for material or optimum experimental design using quantum computing algorithms available in quantum computing frameworks, such as Qiskit[28]. Therefore, further studies can bring new perspectives in the field of QAL or Quantum Bayesian optimization, mostly their search efficiency due to: initial data size and distribution; role of uncertainty quantification in the decision making – e.g.: K-fold cross-validation and non-parametric bootstrap[29] together with the number of resamplings[8]; type of acquisition function in the search – e.g.: expected improvement (EI); lower / upper confidence bound[27]: LCB / UCB; probability of improvement[27]: PI); role of the quantum circuits encoding in the search and the number of qubits; influence of the quantum kernel functions in the search or regression quality.

Considering that most of this plethora of possibilities used to design robust and efficient "AI agents" were implemented in the MLChem4D software, in this work we have chosen to study the efficiency of QAL for experimental design for three perovskite oxide databases using the QSVR algorithm, the EI acquisition function for decision making and the K-CV with K=5 for uncertainty quantification. In addition, quantum kernels were obtained using the following quantum feature encoding: XFeatureMap (X = Z, ZZ and Pauli)[30], available in Qiskit; HighDim and YZ_CX, available in sQUlearn[23]. Note that X = Z prepares an unentangled initial state whereas for X = ZZ and Pauli an entangled starting state is produced. We examine the effect of such choices on the efficiency of the QAL process for the perovskites. In addition, the HighDim and YZ_CX quantum circuits are entangled.

The data set used in this QAL study for: (I) the piezoelectric coefficient ($d_{33}$) maximization of modified single perovskites $Ba_{(1-x)}A_xTi_{(1-y)}B_yO_3$ (A = Ca, Sr, Cd; B = Zr, Sn) was given by Ref[8]; (II) the band-gap minimization of double-perovskites was obtained from Ref[31]; (III) the energy storage density maximization of $Ba_{(1-x)}A_xTi_{(1-y)}B_yO_3$ A = Ca, Sr,

Cd; B = Zr, Sn, Hf) modified single perovskites was acquired from Refs[8,12]. System (IV) is about applying QAL for automatic structural determination jointly with spin multiplicities of an $Si_{11}$ nanoparticle doped with Al ($3Al@Si_{11}$). Then, for each Al distribution, the spin multiplicities 2, 4 and 6 also changed during the QAL cycles, allowing a powerful methodology to optimize, at once, both structures and spin multiplicities[32].

To the best of our knowledge, this work is amongst the first applications of QAL[33] and the first in the area of materials design, using quantum regression algorithms. While writing this paper we became aware of the QAL work reported in Ref[33] for classification problems. Our results revealed that the QAL method improved the searches in most cases, but not all, seemingly correlated with the "roughness" of the data. We emphasize that the study has been carried out on quantum emulators in noise-free mode, so the results should be taken in the context of beyond-NISQ machines. Moreover, QAL is a hybrid approach, with the data-encoding and the definition of the kernel being quantum parts of the protocol, while a classical SVR or GP algorithm is used for the regression, with a quantum kernel.

## *Results*

The results for the proposed quantum active learning (QAL) method were obtained for three different perovskite problems: (I) piezoelectric coefficient maximization of modified single perovskites $Ba_{(1-x)}A_xTi_{(1-y)}B_yO_3$ (A = Ca, Sr, Cd; B = Zr, Sn); (II) band-gap minimization of modified double perovskites $AA'BB'O_6$ and (III) maximization of energy storage density of $Ba_{(1-x)}A_xTi_{(1-y)}B_yO_3$ (A = Ca, Sr, Cd; B = Zr, Sn, Hf). Although the search in $Ba_{(1-x)}A_xTi_{(1-y)}B_yO_3$ to find perovskites with optimum piezoelectric coefficients and energy storage densities involves similar elements except for Hf in the B site, their elemental composition is different. The former has a database with 74 and the latter a database with 242 $Ba_{(1-x)}A_xTi_{(1-y)}B_yO_3$ perovskites[8]. The band-gap database has 54 $AA'BB'O_6$ double perovskites with promising water splitting catalytic properties. The QAL efficiency was compared to classical active learning (AL) for the aforementioned systems using SVR with a radial basis function (RBF) kernel[25,34] (more information can be found in section 2 of SI). Twenty independent runs were taken into account. For each one a different distribution of the initial data set was obtained randomly. This is fundamental to gaining insight into the role of different data distributions for the quality of the agent – the decision making process by the QSVR algorithm and its uncertainty – since, in practice, data distribution is not fixed when designing and doing experiments. Also, the perovskite configurations with properties closer to the optimum one were removed from the random initial data set. Thus, the aim was to test the QAL algorithm in a challenging material discovery scenario – small data set with perovskite properties far away from the optimum one with a different properties distribution. In all studies the EI acquisition function was used for decision making. QAL uses QSVR regression with the following quantum circuits (QC) encoding: XFeatureMap (X = Z, ZZ and Pauli)[30]. In particular, the PauliFeatureMap and ZZFeatureMap were used with the qubits with full entanglement. On the contrary, the ZFeatureMap QC qubits had no entanglement. The QCs are used to obtain the fidelity quantum kernel. The results were compared to the classical AL where the SVR regression with EI was used. The radial basis function (RBF) kernel was used in classical SVR. In this study, the agent selects one new material ($N_{selected} = 1$, Fig. 5 in the Methods section) in each AL or QAL cycle. Further information about statistical regression as well as classical and quantum kernels can be found in SI and in the Methods section.

An additional application of the proposed QAL method is for system IV: automatic structural determination jointly with spin multiplicities of an $Si_{11}$ nanoparticle doped with three Al ($3Al@Si_{11}$). For each Al distribution, the spin multiplicities (SM = 2, 4 and 6) also changed during the QAL or quantum Bayesian optimization[26] (QBO) search. The data base was obtained from Ref[32] and has 165 structures or homotops of $3Al@Si_{11}$ (and three spin multiplicities for each: 2, 4 and 6) calculated by DFT using deMon2k[35]. Ten independent runs were taken into account and for each one a different distribution of the initial $3Al@Si_{11}$ structures was obtained randomly to define the initial random data base – composed with a total of 20 data, $N = 20$: structures with SMs –, making sure the lowest energy structures are far away from the global minimum – which is a doublet (SM = 2). In this study, for each AL or QAL iteration, the agent[14] proposed a batch of five ($N_{selected} = 5$, Fig. 5 in Methods section) possible stable structures together with the spin multiplicities from the pool of $3Al@Si_{11}$ homotop candidates (in the unexplored space), which are sent to be validated with a DFT local minimization. The number of AL or QAL cycles is: $N_{cycles} = 60$. More details about the $3Al@Si_{11}$ system and the classical AL method for structural determination using classical AL can be found in Refs[21,36-39].

## *Application to piezoelectric ($d_{33}$) coefficient (pC/N) maximization of $Ba_{(1-x)}A_xTi_{(1-y)}B_yO_3$ (A = Ca, Sr, Cd; B = Zr, Sn) perovskites*

In this study seven features were used to describe the *$Ba_{(1-x)}A_xTi_{(1-y)}B_yO_3$ (A = Ca, Sr, Cd; B = Zr, Sn)* perovskites. Thus, no dimensionality reduction was used and 7 qubits were considered to build the feature maps XFeatureMap (X = Z, ZZ and Pauli) – data encoding. The features were scaled using the standard scaler method implemented in Scikit-learn[40]. The QAL started with 22 random data where all perovskites among them had the piezoelectric coefficient smaller than or equal to 300 pC/N.

Fig. 1A shows the results of the average piezoelectric ($d_{33}$) coefficient (pC/N) as a function of the number of experiments indicated by the QAL agent for 20 independent runs. Different QSVR – due to different QC encoding to obtain the kernel – using EI (QSVR-EI) as the acquisition function were explored in QAL. The results were compared to classical SVR (SVR-EI) – classical AL. As the quantum kernel (used in the SVR) depends on the quantum circuit (QC) encoding (e.g.: XFeatureMap, X=Z, ZZ and Pauli), different results for the $Ba_{(1-x)}A_xTi_{(1-y)}B_yO_3$ *(A = Ca, Sr, Cd; B = Zr, Sn)* searches were found. In particular the PauliFeatureMap and ZZFeatureMap QC used full quantum entanglement of the qubits. On the contrary, the ZFeatureMap QC has no entanglement.

The results in Fig. 1A show a substantial improvement of the QAL-EI for all QC explored in this search for modified $Ba_{(1-x)}A_xTi_{(1-y)}B_yO_3$ *(A = Ca, Sr, Cd; B = Zr, Sn)* perovskites with maximum piezoelectric coefficient ($d_{33}$). The QSVR-EI with ZZFeatureMap performed better (red curve). These results are similar to the QSVR-EI one with PauliFeatureMap (green curve). The QSVR-EI with ZFeatureMap underperformed the QSVR-EI with ZZFeatureMap and PauliFeatureMap (with full entanglement in the quantum circuit qubits). In all cases, the classical AL (using SVR-EI with RBF kernel) yielded average piezoelectric coefficients smaller than those obtained by QAL.

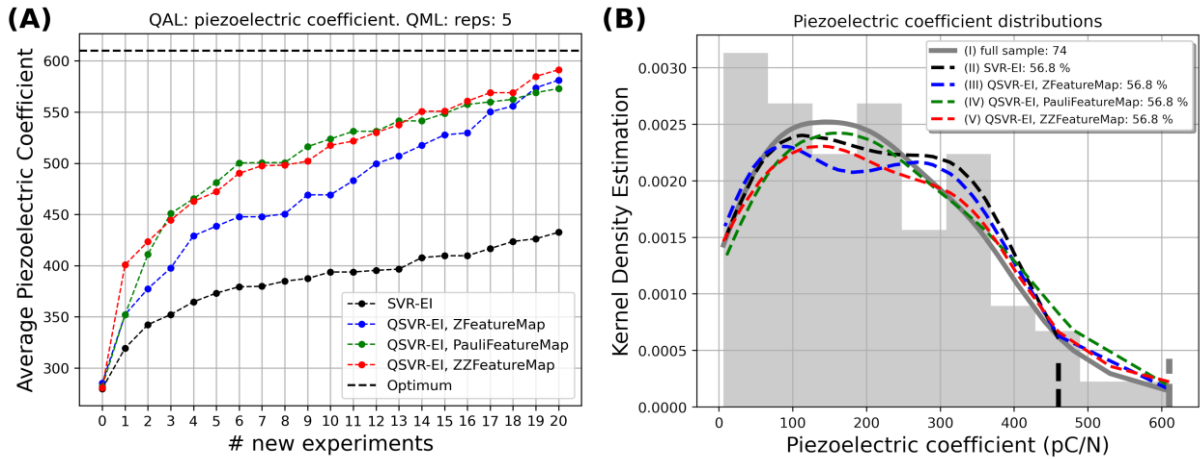

*Figure 1- (A) Average energy piezoelectric (d33) coefficient (pC/N) of $Ba_{(1-x)}A_xTi_{(1-y)}B_yO_3$ (A = Ca, Sr, Cd; B = Zr, Sn) perovskites obtained by 20 independent QAL or AL runs as a function of new experiments (with $N_{cycles} = 20$). SVR: Classical Support Vector Regression with Radial Basis Function (RBF) kernel. QSVR: Quantum Support Vector Regression with XFeatureMap (X=Z, ZZ, Pauli) and Fidelity Quantum Kernel (FQK). EI: Expected Improvement. QML: quantum machine learning; "reps": number of times the quantum circuit is decomposed. (B) Kernel density estimation of piezoelectric coefficient (pC/N) for different $Ba_{(1-x)}A_xTi_{(1-y)}B_yO_3$ (A = Ca, Sr, Cd; B = Zr, Sn) perovskite samples: (I) full sample with 74 data (gray); (II) sample obtained with AL with SVR-EI (black); (III) sample obtained with QAL with QSVR-EI with ZFeatureMap (blue); (IV) sample obtained with QAL with QSVR-EI with PauliFeatureMap (green); (V) sample obtained with QAL with QSVR-EI with ZZFeatureMap (red). The gray and black vertical bar are the maximum piezoelectric coefficient found in the full sample and in SVR-EI, respectively. The percentage of perovskites sampled by QAL or AL is shown (taking the full sample as a reference). One independent run was used in the analysis.*

Looking at Tab. S1 (supporting information, SI), we can see the training and testing MAE ratio (in pC/N) for ~42 data is: 3.07 for ZFeatureMap; 2.26 for ZZFeatureMap; 3.00 for PauliFeatureMap and 3.26 for SVR (with RBF kernel). Values closer to 1.00 in this ratio indicate better regression – less over fitting. This seems to provide evidence for the better piezoelectric coefficient maximization for QAL as compared with classical AL, since the regression quality is better in QSVR. Of course, as regression is dependent on data size and data distribution, this ratio in other scenarios might change.

Fig. 1B presents the histogram and the kernel density estimation[41] (KDE) of the piezoelectric coefficient distribution for the full sample (with 74 perovskites) and the sample obtained by new experiments indicated by the QAL. This analysis considered just one run of the QAL and AL methods using MLChem4D. The gray and black vertical bars are the maximum piezoelectric coefficient found in the full sample and in SVR-EI, respectively. Notice that the QAL methods using QSVR with different data encoding to obtain the FQK were able to find the perovskite with the highest piezoelectric coefficient (gray vertical black). The classical AL with SVR failed in finding the perovskite with the highest piezoelectric coefficient (black vertical bar). In general, the QAL methods found perovskites with piezoelectric coefficients whose distributions are similar to the full sample. However, the purpose of QAL is not modelling the full distribution; but, to find the optimum material with few new experiments.

## *Application to band-gap minimization of double AA'BB'O$_6$ perovskites*

The double AA'BB'O$_6$ perovskites were described by 64 features based on the elemental properties and the perovskite compositions – Ref[31] shows how the features were obtained and more information can be found in SI. In this case, dimensionality reduction using PCA[42] was used to reduce it to 8 components. In QML this resulted in 8 qubits, which were used to obtain the feature maps XFeatureMap (X = Z, ZZ and Pauli) – for data encoding. The QAL study started with 10 random data, guaranteeing that all perovskites among the 10 had a band-gap equal to or greater than 2.0 eV.

In the search for double perovskites (AA'BB'O$_6$) with small band-gaps, the average band gaps obtained for 20 independent QAL and AL runs are shown in Fig. 2A. The XFeatureMap (X = Z, ZZ and Pauli) QCs were used to obtain the fidelity quantum kernel, used in QSVR. The QAL-EI with ZFeatureMap encoding found, in most of the QAL cycles, double-perovskites with smaller band gaps (blue curve). After cycle 17 the QAL-EI with ZZFeatureMap (red curve) presented a deeper decline of the average band gaps, when compared to the QAL-EI with ZFeatureMap and PauliFeatureMap. Finally, when comparing the QAL with classical AL, we see that all QSVR-EI methods outperformed the SVR-EI method for band-gap search until the very end of the 20 new experiments, by which point the classical SVR found on average (in 20 independent runs) the global minimum more times, making the best QAL (ZZFeatureMap) get an average band-gap slightly higher.

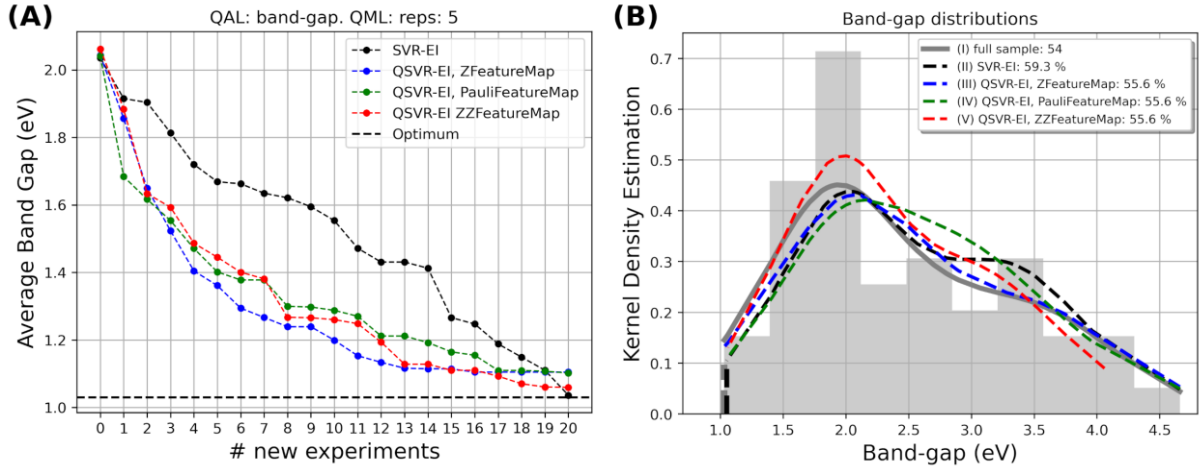

*Figure 2- (A) Average energy band-gap (eV) of doped $A_2BB'O_6$ perovskites obtained by 20 independent QAL or AL runs as a function of new experiments with $N_{cycles} = 20$. SVR: Classical Support Vector Regression with Radial Basis Function (RBF) kernel. QSVR: Quantum Support Vector Regression with XFeatureMap (X=Z, ZZ, Pauli) and Fidelity Quantum Kernel (FQK). EI: Expected Improvement. QML: Quantum Machine Learning. "reps": number of times the quantum circuit is decomposed. (B) Kernel density estimation of band-gap (eV) for different $A_2BB'O_6$ perovskites samples: (I) full sample with 54 data (gray); (II) sample obtained with AL with SVR-EI (black); (III) sample obtained with QAL with QSVR-EI with ZFeatureMap (blue); (IV) sample obtained with QAL with QSVR-EI with PauliFeatureMap (green); (V) sample obtained with QAL with QSVR-EI with ZZFeatureMap (red). The gray and black vertical bar are the minimum band-gap found in the full sample and in SVR-EI, respectively. The percentage of perovskites sampled by QAL or AL is shown (taking the full sample as a reference). One independent run was used in the analysis.*

From Tab. S1 (SI) we can obtain the training and testing MAE ratio (in eV) for ~20 data: 2.33 for ZFeatureMap; 4.19 for ZZFeatureMap; 1.31 for PauliFeatureMap and 2.23 for SVR (with RBF kernel). When this ratio is ~1.00, better is the regression or less over fitting. The situation is ambiguous in this case, no clear distinction can be made between the quantum and classical prediction protocols. Of note, as regression depends on data size and data distribution, this ratio can change in other scenarios.

The KDE for the band-gap distribution for the full double perovskites sample (54 AA'BB'O$_6$ materials) and the ones obtained by QAL and AL is presented in Fig. 2B. Also, the histogram for the full sample is shown. One independent run of the QAL and AL methods was considered in the analysis. The QAL methods using QSVR with different feature maps to obtain the FQK were able to find the perovskite with the smallest band-gap (gray vertical line). For this independent run, the classical AL with SVR found a AA'BB'O$_6$ perovskite with a band-gap a bit above the optimum one, as seen in the black vertical line, Fig. 2B. The QAL methods found AA'BB'O$_6$ perovskites with band-gaps whose distributions are similar to the full sample. The agent in QAL doesn't aim to find materials whose band-gaps follow the full distribution; but, to find with few experiments the next ones to be observed in order to optimize the desired property.

## *Application to energy storage density maximization of Ba$_{(1-x)}$A$_x$Ti$_{(1-y)}$B$_y$O$_3$ (A = Ca, Sr, Cd; B = Zr, Sn, Hf) perovskites*

Here, 34 features based on the elemental properties and the perovskites composition (as explained in details in Ref[20] and in SI) were used to describe the Ba$_{(1-x)}$A$_x$Ti$_{(1-y)}$B$_y$O$_3$ *(A = Ca, Sr, Cd; B = Zr, Sn, Hf)* perovskites. Then, dimensionality reduction using PCA was applied to reduce the features to 8 components, which were subsequently mapped onto 8 qubits in the QC. These qubits were used to obtain the data encoding XFeatureMap (X = Z and Pauli) – feature maps (ZZ was not examined because the results for these two feature maps were quite similar for the previous two cases and ZZ is computationally demanding for this data size). The QAL study started with 73 random data, all 73 perovskites had their energy storage densities equal to or smaller than 65 mJ/cm$^3$.

The efficiency of the QAL search in finding modified perovskites (Ba$_{(1-x)}$A$_x$Ti$_{(1-y)}$B$_y$O$_3$) with maximum energy storage density was done by emulating a laboratory condition by starting with 73 Ba$_{(1-x)}$A$_x$Ti$_{(1-y)}$B$_y$O$_3$ *(A = Ca, Sr, Cd; B = Zr, Sn, Hf)* random perovskites from an experimental database of 242 perovskites[8]. During the random process, these 73 perovskites were randomly selected by constraining their energy storage density to be below or equal to 65 mJ/cm³ – a challenging optimization scenario. In each AL cycle or iteration ($k$) $N_{selected}^{k+1}$ = 1. This indicates the number of selections by the acquisition function or agent in the unexplored space ($N_{virtual}^{k}$) for new observations.

Fig. 3A shows the results of the QAL method (as implemented in MLChem4d) in maximizing the energy storage density (mJ/cm³) of modified perovskite oxides ($Ba_{(1-x)}A_xTi_{(1-y)}B_yO_3$). Until 15 QAL cycles or new experiments the QAL method using QSVR-EI with ZFeatureMap (no qubit entanglement) presented the highest average energy storage density for 20 independent runs. In most of these cycles or new experiments the QSVR-EI with PauliFeatureMap-circ (where the circular entanglement of the qubits in the QC was considered) presented a similar performance when compared to classical AL (SVR-EI). When considering the QAL with QSVR-EI with PauliFeatureMap-full (full entanglement) the performance, in most of the cycles, was a bit lower than the other QAL methods.

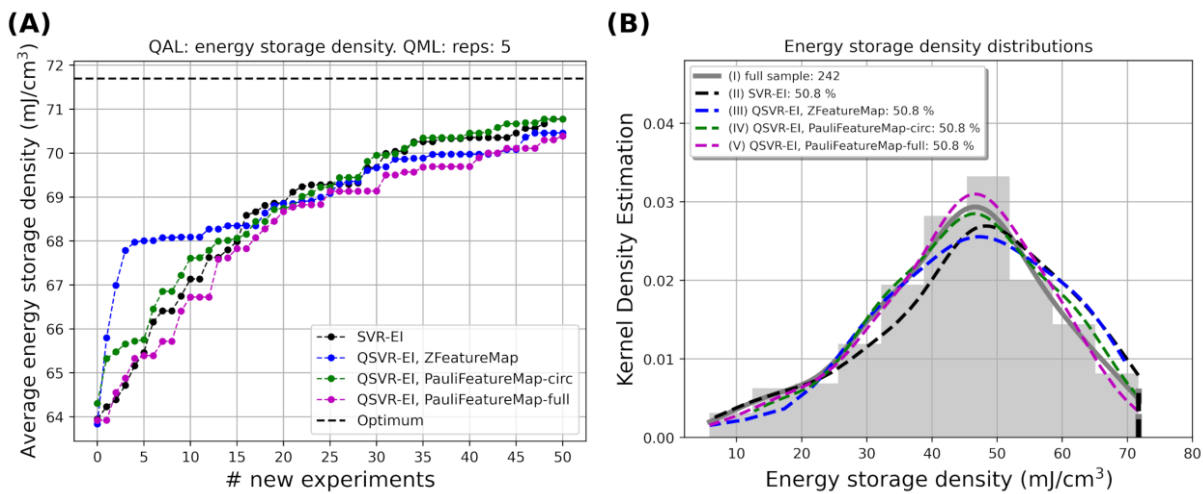

*Figure 3- (A) Average energy storage density (mJ/cm₃) of $Ba_{(1-x)}A_xTi_{(1-y)}B_yO_3$ (A = Ca, Sr, Cd; B = Zr, Sn, Hf) perovskites obtained by 20 independent QAL or AL runs as a function of new experiments (with $N_{cycles}$ = 50). SVR: Classical Support Vector Regression with Radial Basis Function (RBF) kernel. QSVR: Quantum Support Vector Regression with XFeatureMap (X=Z, Pauli) and Fidelity Quantum Kernel. EI: Expected Improvement. QML: Quantum Machine Learning. "reps": number of times the quantum circuit is decomposed. (B) Kernel density estimation of energy storage density (mJ/cm³) distributions for different $Ba_{(1-x)}A_xTi_{(1-y)}B_yO_3$ (A = Ca, Sr, Cd; B = Zr, Sn, Hf) samples: (I) full sample with 242 data (gray); (II) sample obtained with AL with SVR-EI (black); (III) sample obtained with QAL with QSVR-EI with ZFeatureMap*

*(blue); (IV) sample obtained with QAL with QSVR-EI with PauliFeatureMap-circ (green); (V) sample obtained with QAL with QSVR-EI with PauliFeatureMap-full (pink). The gray and black vertical bar are the maximum energy storage density found in the full sample and in the AL with SVR-EI, respectively. The percentage of perovskites sampled by QAL or AL is shown (taking the full sample as a reference). One independent run was used in the analysis.*

From Tab. S1 (SI) the training and testing MAE ratio (in $mJ/cm^3$) for ~113 data are: 2.42 for ZFeatureMap; 6.57 for PauliFeatureMap and 1.21 for SVR (with RBF kernel). The MAE ratio ~1.00 indicates a regression model with less over fitting. In this case, the classical algorithm does better than either quantum approach, on this metric. This analysis should be seen considering that regression models are dependent on data size and data distribution. Thus, these ratios can change in other scenarios.

The KDE analysis for the energy storage density of the new $Ba_{(1-x)}A_xTi_{(1-y)}B_yO_3$ perovskites indicated by the QAL agent is shown in Fig. 3B together with the histogram for the full sample – which has 242 $Ba_{(1-x)}A_xTi_{(1-y)}B_yO_3$ perovskites. The QAL and AL methods were able to find the perovskite with the smallest band-gap (gray and black vertical lines). The QAL found $Ba_{(1-x)}A_xTi_{(1-y)}B_yO_3$ perovskites with energy density storages with distributions similar to the full sample. These KDEs resemble a normal distribution. An independent run of the QAL and AL methods was considered to obtain Fig. 3B.

## *Application to structural determination and optimum spin multiplicity of $3Al@Si_{11}$ nanoparticles*

Also, we have proposed a QAL or QBO[26] for automatic structural determination of doped nanoparticles together with the search for the optimum spin multiplicity on-the-fly. This was implemented in the QMLMaterial[21] software using the sQUlearn[23] quantum computing library and was applied for the $3Al@Si_{11}$ doped nanoparticle. In addition to the change of the position of the three Al doped in $Si_{11}$ ($3Al@Si_{11}$) the possible spin multiplicities (SM = 2, 4 and 6) were allowed to change for each possible doped structure in the QAL framework. More details about this method using classical AL can be found in Ref[32]. In this study, the agent proposed a batch of five ($N_{selected}$ = 5, Fig. 5 in the Methods section) possible stable structures together with the spin multiplicities from the pool of

3Al@Si$_{11}$ homotops to be validated with DFT. The initial population (N) was 20. More details can be found in Ref[32].

The many body tensor representation (MBTR) method, available in the DScribe[43] library and interfaced in QMLMaterial, was used to describe the 3Al@Si$_{11}$ homotop structures. Also, a SM descriptor was used to represent the multiplicities, as described in SI. The two descriptors, MBTR, for structures, and spin multiplicity, were concatenated. This joint descriptor has its dimensionality reduced using PCA, whose final dimension was 4, aiming to use just 4 qubits for data encoding using the HighDim and YZ_CX circuits (Figs. S5 and S6). Tab. S2 (in SI) presents the optimum hyperparameters found and the statistics of both GPR and QGPR models for different data sizes during the AL and QAL iterations.

Fig 4A shows the average total energy (in Hartree) of 3Al@Si$_{11}$ homotops (for 10 independent runs) as a function of the number of calculations (DFT local optimizations using the deMon2k program). Different QAL studies using QGPR were considered by changing: the quantum kernel (PQK, Eq. 6 and FQK, Eq. 5 – Methods section); quantum circuits HighDim and YZ_CX as defined in the sQUlearn[23] library (Figs. S5 and S6); acquisition function: EI and exploitation (prediction only). The QAL performance in minimizing the energy with optimum spin multiplicities for 3Al@Si$_{11}$ can be compared with the classical AL (black curves in Fig. 4A). Quite surprisingly – as opposed to classical GPR whose kernel is a combination of DotProduct and Whitekernel, Eq 2 – the QGPR models presented small uncertainty values, independent of the quantum circuit, quantum kernel and data size. This explains why exploitation yielded better performance than EI. As discussed in the Methods section, from Eq. 8 we can see that by having T higher (defined: $T = f_{min} - \mu(x^{(j)})$) and $\sigma(x)$ lower, EI ~ T, which means we are exploiting the already known region of the surface.

In classical AL, the tradeoff between exploitation, $\mu(x^{(j)})$, and exploration, $\sigma(X)$, using EI resulted in better performance (black curve: "GPR-EI") when compared to exploitation (black curve: "GPR-Exploit."). On the contrary, the QAL presented better performance when the decision making was based on pure exploitation (without uncertainty for exploration). For example, when considering the HighDim and PQK kernel, "QGPR-Exploit." presented better results – pink curve in Fig. 4A – when compared to "QGPR-EI" (blue curve), when EI was used for decision making.

The "QGPR-Exploit." with HighDim quantum circuit for data encoding and quantum kernel PQK performed better in obtaining the putative global minimum (GM) structure with the optimum SM (SM=2, a doublet). This outperformed the classical AL with GPR-EI.

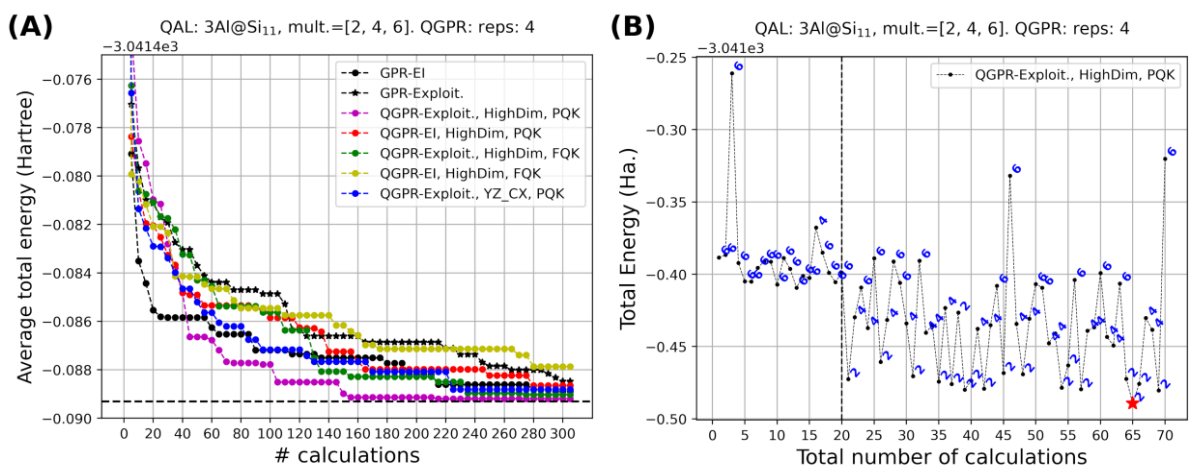

*Figure 4- (A) Average total energy obtained by DFT (in Hartree) for $3Al@Si_{11}$ obtained by 10 independent QAL or AL runs as a function of new calculations (with $N_{cycles} = 60$ and $N_{selected} = 5$). GPR: Classical Gaussian Process Regression with DotProduct and WhiteKernel combination. QGPR: Quantum Gaussian Process Regression with XFeatureMap (X=HighDim, YZ_CX) and Projected Quantum Kernel (PQK) or Fidelity Quantum Kernel (FQK). "Exploit.": pure exploitation. EI: Expected Improvement. "reps": number of times the quantum circuit is decomposed. (B) Total Energy (in Hartree) for different spin multiplicities (2, 4, 6), SMs, as a function of the total number of local optimizations for one run. The left side of the vertical line shows the structures and SMs energies of the 20 initial data. The right side presents those obtained from the QAL method by QGPR, with EI, HighDim quantum circuit and PQK. The global minimum is represented by a red star (SM = 2). The MBTR descriptor was used for the $3Al@Si_{11}$ structures together with a SM descriptor to represent the multiplicities, as shown in SI.*

Fig 4B shows the total energy (in Hartree) as a function of the number of calculations for a particular QAL run using QGPR with exploitation and the HighDim quantum circuit with PQK. On the left side of the vertical line are the 20 initial data obtained randomly; on the right side are the energies obtained from the QAL method by using QGPR with spin multiplicities (SM): 2, 4 and 6. The red star represents the GM, which is a doublet. For this particular run, the QAL method, as implemented in QMLMaterial, required 65 new local optimizations (a total of 85) to find the putative GM. Indeed, the QAL with SM is able to probe different structure surfaces and SM at the same time, providing "spin jumping" between potential energy surfaces during the QAL cycles.

## *Discussion*

Artificial intelligence (AI) methods based on active learning (AL) have been shown to be efficient for decision-making in situations where small datasets of materials are available. The meeting of AI and quantum computing technologies is already a reality[4] and *Quantum Artificial Intelligence* (QAI) is emerging as a new field. In this work, a QAI method based on active learning (QAL) for experimental or material design was developed and implemented in: MLChem4D, targeting experimental or material design; QMLMaterial, aimed at automatic structural determination of nanoparticles (e.g. doped ones) using DFT to build the potential energy surface. The QAL method was applied to single and double modified perovskite oxides (using MLChem4D[20]) as well as to the structural determination together with the spin multiplicity of $3Al@Si_{11}$ (using QMLMaterial[21]).

The results of the proposed QAL for the single and double modified perovskites (i.e.: maximization of piezoelectric coefficients, Fig. 1; minimization of band-gaps, Fig. 2; and maximization of energy storage densities, Fig. 3) indicate that using QSVR trained on-the-fly during the QAL cycles with the fidelity quantum kernel obtained by different quantum circuits (e.g.: XFeatureMap, X=Z, ZZ and Pauli) outperformed the results obtained from classical AL using SVR with the RBF kernel for maximization of piezoelectric coefficients (Fig. 1A) and for minimization of band-gaps (Fig. 2A). However, for the maximization of the energy density storage, the QAL application outperformed the AL (Fig. 3A) just in the first few QAL new experiments; after that the QAL methods with different quantum circuits yielded similar performance when compared to classical AL.

Seeking insight into those situations that offer some quantum advantage and those that don't, we offer the following preliminary, and perforce somewhat conjectural, analysis. Comparing the KDE in Figures 1B, 2B and 3B we see, first (Fig. 1B) a distribution that descends more or less smoothly from the origin, then (Fig. 2B) a highly skewed distribution with a maximum around 2eV and finally (Fig. 3B) a slightly skewed distribution that is more-or-less normal. The relative performance of our QAL protocol, compared to classical AL, decreases in the same order, with Fig. 1A showing the clearest quantum improvement, Fig. 2A requiring a more nuanced analysis: quantum better at the early stages but classical catching up around the 20 new calculation mark, and Fig. 3A showing mostly similar performance for classical and quantum approaches (with just some early advantage for, intriguingly, the quantum approach without entanglement). We note in Ref[44] a somewhat similar conclusion in a quantum reservoir-computing study of time sequence data. They found a quantum advantage for high-frequency (rough?) data but none for low-frequency data. Clearly, more work will be required to see if this trend, that quantum improvement might be found for "rough" data sets might generalize.

The QAL or QBO using QGPR for automatic structural determination has shown to be efficient, when compared to classical AL using GPR. The results of QAL for $3Al@Si_{11}$ using the HighDim quantum circuit with 4 qubits and PQK ("QGPR-Exploit., HighDim, PQK", Fig. 4) outperformed all other QAL and AL methods, as highlighted in Fig. 4A. All AL and QAL reported in Fig. 4 used the MBTR descriptor for the structures of $3Al@Si_{11}$ jointly with the spin multiplicities — as described in more details in SI and Ref[32]. The success of "QGPR-Exploit., HighDim, PQK" using just exploitation for decision making (since the uncertainties were small in all QAL studies using QGPR), shows that the quantum circuits, quantum kernels and dimensionality reduction techniques influence the epistemic uncertainty[8,45] (whether this is obtained analytically from GP or resampling techniques, such as CV and BS) and how this affects the agent[14] in making better decisions (i.e.: exploitation and exploration tradeoff) to indicate new materials or structures for further observations, with few data.

The results presented here explore just a small subset of the vast number of applications that will be needed to establish the credentials of QAL. The work brings several perspectives to be explored with quantum kernel algorithms (e.g.: QSVR and QGP) within a QAL workflow – aiming to find improved materials' efficiently. For instance, several QAL "parameters" can be explored to investigate the performance of the agent in finding new or faster discoveries: type of uncertainty quantification besides K-fold CV, e.g. bootstrap (BS)[29]; the influence of the number of K-Fold CV splits as well as of BS in the uncertainty quantification: exploitation and exploration tradeoff; different types of quantum kernels[2] (e.g.: Fidelity and Projected Quantum Kernels); other types of feature maps and quantum entanglement protocols[46]; different acquisition functions, such as the lower or upper confidence bound[27]; the role of the number of qubits required to describe the descriptors and the type of dimensionality reduction[47]. From these: the CV and BS; the XFeatureMap (X = Z, ZZ, Pauli) with several entanglements – as available in Qiskit[28] framework – and HighDim and YZ_CX from sQUlearn[23]; the LCB and UCB acquisition functions are available in MLChem4D[20] as well as in QMLMaterial[21] to investigate the uncharted chemical space aiming at better materials design and discovery, guided by QAL.

There are two essential differences between quantum kernel methods and classical SVMs. On one hand, quantum computers have the potential to be exponentially more powerful than classical computers, allowing quantum feature maps to be more expressive than classical ones. This increased expressivity means that quantum feature maps can represent complex data patterns and relationships more effectively, capturing intricate structures that classical feature maps might miss, potentially enhancing performance in pattern recognition, classification and regression tasks. On the other hand, quantum kernels are subject to small additive noise from the estimation process, whereas classical kernels can be computed exactly. Therefore, in real scenarios, demonstrating a quantum advantage with the QML algorithms requires leveraging the expressivity of quantum feature maps while ensuring robustness against finite sampling noise. However, it should be noted that the current study has been conducted under ideal conditions, with no noise considered. When comparing the expressivity of the employed feature maps, the ZFeatureMap encodes data using rotations around the Z-axis of the qubit's Bloch sphere. While simple and efficient, its expressivity is limited because it only leverages a single axis of rotation. The ZZFeatureMap incorporates interactions between pairs of qubits, adding an additional layer of complexity by using entangling gates. This allows the model to capture correlations between data points, enhancing its expressivity compared to the ZFeatureMap and making it more capable of representing complex patterns and relationships in the data. Lastly, the PauliFeatureMap employs rotations around multiple axes (X, Y, and Z) and includes entangling gates, offering a comprehensive approach to data encoding. With its multi-axis rotations and entanglement, the PauliFeatureMap has the highest expressivity among the three. It can capture intricate data structures and complex relationships, making it highly suitable for challenging pattern recognition and classification tasks. In this work, the quantum feature maps feed into a quantum kernel calculation, then a classical SVR or GP process. Future work could focus on quantum algorithms, such as the variational quantum classifier/regressor; a quantum classifier algorithm was used in earlier work by some of us to classify materials as perovskites[48].

Finally, the MLChem4D and QMLMaterial programs (written in Python3.x) now supports quantum machine learning algorithms that can be explored with different quantum circuits for data encoding. These QAI programs are currently under development with the purpose of providing a general method for experimentalists or theoreticians to aid chemists, physicists or materials engineers in finding or understanding the nature of materials with the proper elements and compositions to be synthesized and to have their properties evaluated, aiming to improve their properties with as few experiments as possible – from experiment or from computational modelling. This QAI technology has the potential of finding optimum solutions, within the chemical space, in materials science and in other branches of chemistry.

## *Methods*

### *Active learning*

The QAL method presented in this work uses supervised QML algorithms (quantum regression) and their epistemic uncertainties to make decisions for the next conditions in further experiments[27]. The QML regression predictions (see section 1, 2 and 3 of supporting information, SI, for more details) are improved in each cycle since more data is obtained, increasing the probability of finding the optimum material with the property of interest improved or the optimum structure of a nanoparticle. The sequence of the QAL method developed in the present work is shown in Fig. 5:

(A) *Data on materials*: a database with *N* observed materials (synthesized or computed or from legacy data) is set up. At this moment, the descriptor for composition of perovskites or isomers of the nanoparticle is provided (i) for the observed materials

(N) – $x^{(i)} = (x_1^{(i)}, ..., x_k^{(i)})$, $\forall$ $i$ = 1,..., N – and for the unexplored materials or virtual space ($N_{virtual}^k$) – $x^{(j)} = (x_1^{(j)}, ..., x_k^{(j)})$, $\forall j$ = 1,..., $N_{virtual}^k$. Details about the atomic based descriptors for ABO$_3$ and AA'BB'O$_6$ perovskites as well as for structural and spin multiplicities of doped nanoparticles can be found in the SI.

(B) Quantum machine learning or *decision-making (the agent)*: a QML regression model (ii) is obtained for the N observed materials and the uncertainty is computed (iii). From that, the agent – based on decision-making (iv) like the EI (Eq. 8) – informs *next experimental condition* or new material (in the unexplored space: $N_{virtual}^k$), $N_{selected}^{k+1}$, to be measured in the laboratory.

(C) *New experiments or computations*: $N_{selected}^{k+1}$ *new experiments* are realized and the property of the materials evaluated, increasing the database: ($N$ = $N$ + $N_{selected}^{k+1}$). If the new material with its property is not yet satisfactory, the algorithm returns to step (A) with the new measurement added to the initial database: $N$ = $N$ + $N_{selected}^{k+1}$. New QML regression models and uncertainties are obtained (step B). The cycles continues ($k$) until the property of interest is improved or the budget has been exhausted.

The boxes from (i) to (iv) in Fig. 5 highlight some of the capabilities available in MLChem4D for optimum experimental design and in QMLMaterial for automatic structural determination using QML. In addition, MLChem4D and QMLMaterial are general purpose codes capable of handling different initial data sizes (N) and selections ($N_{selected}^{k+1}$) of the next materials in $N_{virtual}^k$ to be observed in the laboratory. These features allow the use of previous legacy knowledge – from the researchers or even from literature data – of the material to be designed to be incorporated in the initial database, besides giving flexibility to explore different options for the experimental design or automatic structural determination of doped nanoparticles.

More methodological information about statistical regression, classical and quantum kernels can be found in SI.

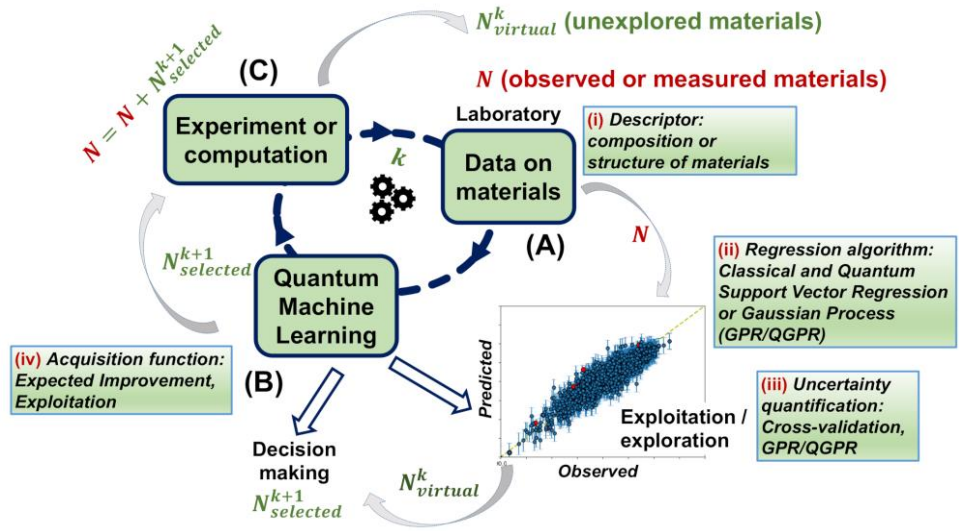

*Figure 5- Quantum Artificial intelligence (AI) workflow based on quantum active learning (QAL) for optimum experimental design. This is implemented in MLChem4D and QMLMaterial software using Qiskit and sQUlearn frameworks, respectively.*

## *Classical kernel Functions*

Kernel based methods for regression – such as SVR and GPR – are widely employed in supervised learning. They require a prior's covariance that needs to be specified by passing a kernel object. In this work, for classical SVR, we used the radial basis function (RBF) kernel, as implemented in scikit-learn library[40] for classical ML. The RBF is given by Eq. 1.

$$k_{RBF}(x_i, x_j) = exp\left(-\frac{d(x_i,x_j)^2}{2l^2}\right) \quad (1)$$

In Eq. 1, $x_i$ and $x_j$ are vectors that describe the perovskites (the descriptor) *i* and *j*. $d(x_i, x_j)^2$ is the square of the difference between $x_i$ and $x_j$. From that, $k_{RBF}$ is a matrix whose dimension is equal to the number of perovskites configuration in the database or in the training model. This is parameterized by the length-scale parameter (*l*).

Another classical kernel explored in this work – for the classical AL for the structural and spin multiplicity determination of 3Al@$Si_{11}$ – is shown in Eq. 2. Also, this kernel is based on the scikit-learn[40] library.

$$k_{White+DotProduct}(x_i, x_j) = k_{DotProduct}(x_i, x_j) + k_{WhiteKernel}(x_i, x_j) \quad (2)$$

"DotProduct" is given by Eq. 3, where $\sigma_0^2$ is a parameter that controls the inhomogeneity of the kernel.

$$k_{DotProduct}(x_i, x_j) = \sigma_0^2 + x_i \cdot x_j \quad (3)$$

The "WhiteKernel" in Eq. 2 is for a better estimation of the noise level of the data and is calculated from Eq. 4, where $\nu$ is the noise level parameter (variance).

$$k_{WhiteKernel}(x_i, x_j) = \nu \delta_{ij} \quad (4)$$

## *Quantum kernel functions*

Quantum Kernels (QK) are quantum analogs of classical kernels and allow us to build QML models. The idea is to find patterns by transforming data in a vector form (or features) into a high dimensional feature space, *F*. The data mapping from the original input space H to *F* is done by a feature map $\varphi$: H $\rightarrow$ *F*. The inner products of the features vectors $\varphi(x_i)$ can access the feature space and a function $k_{FQK}$ of two data points or vectors, $x_i, x_j$, is a kernel $k_{FQK}$, Eq. 5. FQK stands for fidelity quantum kernel.

$$k_{FQK}(x_i, x_j) = \langle \varphi(x_i) | \varphi(x_j) \rangle \quad (5)$$

Note that $k_{FQK}(x_i, x_j)$ depends on the feature map, *F*. The dimension of $k_{FQK}(x_i, x_j)$ matrix is equal to the number of perovskite configurations or doped nanoparticle homotops present in the database.

Another QK is the projected quantum kernel (PQK), which is based on k-particle reduced density matrices (k-RDMs) measurements, Eq. 6. This is defined in Eq. 6 and was obtained using the sQUlearn quantum computing framework that is interfaced in QMLMaterial. In Eq. 6, $\gamma$ is a real and positive hyperparameter; $\rho_k(x_i) =$ is the 1-RDM for qubit k: the partial trace of the quantum state $\rho_k(x_i)$ over all qubits except for the k-th one. Eq. 6 resembles the RBF kernel (Eq. 1), but in PQK kernel the vectors (or descriptors) $x_i$ and $x_j$ are encoded in the quantum circuit or feature map (e.g.: Fig. 6 and Fig. S1 to S6 in SI).

$$k_{PQK}(x_i, x_j) = \exp\left(-\gamma \sum_{k,P} \{tr[P\rho_k(x_i)] - tr[P\rho_k(x_j)]\}^2\right) \quad (6)$$

In our kernel-based quantum machine learning, at the first step the data (Fig. 6A) is encoded into a quantum state, Eq. 7 (Fig. 6B).

$$|\psi(x)\rangle = U(x)|0\rangle \quad (7)$$

$U(x)$ is the encoding operator implemented by a quantum circuit (Fig. 6B). This quantum circuit manipulates an initial n-qubit quantum state $|0\rangle$ based on the data $x \in H$. The components $x_i$ of the data vectors $x$ as angles in one or two-qubit rotation are encoded. Using a feature map, in quantum-kernel-based machine learning, to calculate the quantum kernel, each kernel entry $K(x_i, x_j)$ is obtained by executing the circuit $U^+(x_j)U(x_i)$ on input $|0\rangle^{\otimes n}$ in which n is the number of the features. In Fig. 6B we can see an example of a quantum kernel estimation circuit. Then an $n \times n$ symmetric kernel matrix is formed by using the mutual kernel values between all elements $K(x_i, x_j)$ of the dataset: $K(x_i, x_j) = K(x_j, x_i)$.

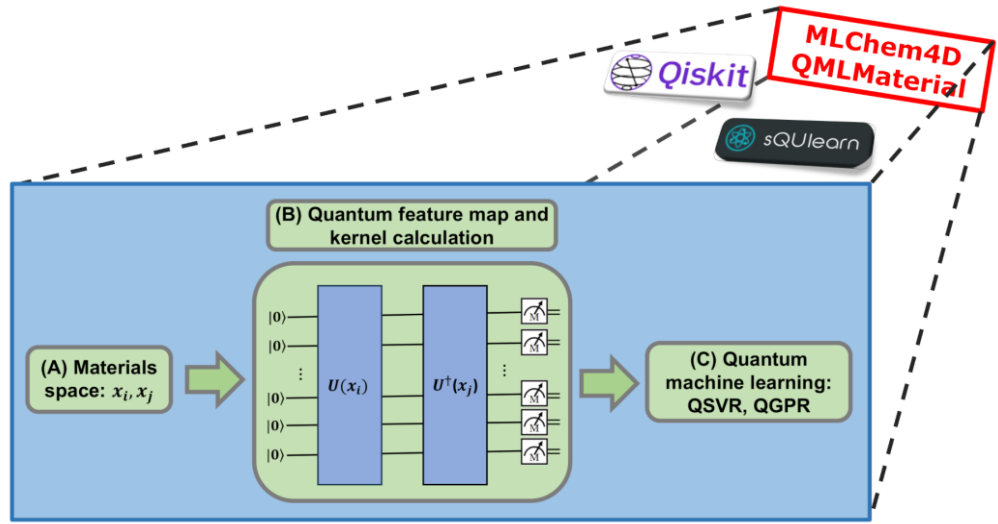

*Figure 6. Quantum machine learning (QML) used in quantum active learning (QAL). (A) Classical data from materials or nanoparticles. (B) Quantum feature map and quantum kernel (e.g. Fidelity Quantum Kernel or Projected Quantum Kernel) calculations. A quantum feature map is used to $(x) \mapsto |\psi(x)\rangle = U(x)|0\rangle$, then each kernel entry $K(x_i, x_j)$ is obtained and the similarity of each pair of data points is estimated by $|\langle 0|^{\otimes n}|U^\dagger(x_j)U(x_i)||0\rangle^{\otimes n}\rangle|^2$. Circuits for the XFeatureMap (X = Z, ZZ and Pauli) used in QSVR (as interfaced in MLChem4D using Qiskit) are*

*given in SI: Figs S1, S2 and S3. Figs S5 and S6 show the HighDim and YZ_CX quantum circuits as implemented in QMLMaterial using sQULearn library. (C) Quantum support vector regression (QSVR) and quantum Gaussian process regression (QGPR).*

Here, each data point (Fig. 6A) is mapped to a quantum state using a quantum feature map (Fig. 6B) and then analyzed in the high-dimensional Hilbert space. The kernel matrix is calculated by measuring the inner product of each pair of data points (Fig. 6B) on a quantum computer, a process known as quantum kernel estimation (QKE). This kernel matrix is subsequently fed into a classical optimizer (Fig. 6C), which efficiently determines the regressor that optimally labels the training data in the feature space.

## *Decision-making*

We consider the descriptors of the materials in the unexplored (virtual) space to be searched: $\boldsymbol{x}^{(j)} = (x_1^{(j)}, \ldots, x_k^{(j)})$, $\forall \; j = 1, \ldots, N_{virtual}^k$. Then, the acquisition or utility function EI[32,37,38,49] for minimization is expressed as:

$$E[I(\boldsymbol{x}^{(j)})] = \left(f_{min} - \mu(\boldsymbol{x}^{(j)})\right) \Phi\left(\frac{f_{min} - \mu(\boldsymbol{x}^{(j)})}{\sigma(\boldsymbol{x}^{(j)})}\right) + \sigma(\boldsymbol{x}^{(j)}) \phi\left(\frac{f_{min} - \mu(\boldsymbol{x}^{(j)})}{\sigma(\boldsymbol{x}^{(j)})}\right) \quad \textbf{(8)}$$

and for maximization[36,50] is:

$$E[I(\boldsymbol{x}^{(j)})] = \left(\mu(\boldsymbol{x}^{(j)}) - f_{max}\right) \Phi\left(\frac{\mu(\boldsymbol{x}^{(j)}) - f_{max}}{\sigma(\boldsymbol{x}^{(j)})}\right) + \sigma(\boldsymbol{x}^{(j)}) \phi\left(\frac{\mu(\boldsymbol{x}^{(j)}) - f_{max}}{\sigma(\boldsymbol{x}^{(j)})}\right) \quad \textbf{(9)}$$

where *j* means the *j-th* material in the virtual space ($N_{virtual}^k$): j = 1, …, $N_{virtual}^k$. The mean and standard deviation are $\mu(\boldsymbol{x}^{(j)})$ and $\sigma(\boldsymbol{x}^{(j)})$, respectively. They can be obtained from CV for SVR or QSVR or from Bayesian statistics using GPR or QGPR, for instance.

The lowest (i.e. formation energy) and the highest (i.e. energy storage density) target property observed so far is $f_{min}$ and $f_{max}$, respectively. $\Phi(\cdot)$ is the cumulative density function and $\phi(\cdot)$ the probability density function. The EI definition can be found in the literature as early as 1978 in the work of Mockus[51] and has been extensively used in global searches, including in the field of materials design.

For the problem of minimization (Eq. 8) as an example, the trade-off between exploitation, min(μ), and exploration, max(σ), is contemplated in the EI. When the target prediction T (defined: $T = f_{min} - \mu(x^{(j)})$) and σ(X) (the uncertainty) are higher, the EI will be higher as well: $EI = T\Phi(\cdot) + \sigma(x)\phi(\cdot)$, Eq. 8. Then, the search takes place farther from the already observed search surface. On the other hand, by having T higher and σ($x$) lower, EI ~ T, which means we are exploiting the already known region of the surface. This is known as exploitation and is more predominant for local minimum searches. Another scenario is when we have T ~ 0 (i.e.: when $f_{min}$ ~ μ) and high σ($x$), thus EI ~ σ($x$), which is known as exploration and it allows jumping to more distant parts of the search landscape. The same analysis is valid for the problem of maximization: Eq. 9.

Note that in QAL, the EI for the unexplored materials or nanoparticle structures – represented by the vector $x^{(j)}$ – to be searched in further experiments is computed using QML models, e.g.: QSVR or QGPR. Therefore, $\mu(x^{(j)})$ and $\sigma(x^{(j)})$ are obtained by taking QML, taking into account the feature map, Fig. 6.

## *Corresponding author*

*Maicon Pierre Lourenço: maiconpl01@gmail.com.

## *Supporting information*

Statistical regression. Quantum circuits used for data encoding. Grid search hyperparameterization for classical (SVR) and quantum (QSVR) machine learning models. Grid search hyperparameterization for classical (GPR) and quantum (QGPR) machine learning models. Quantum machine learning set up. Singe-perovskite descriptor. Double-perovskite descriptor. Spin multiplicity and structural descriptor. Tutorial video explaining how to use the MLChem4D and QMLMaterial programs. References.

## *Statements & declarations*

## *Competing interests*

The authors have no financial interests to disclose.

## *Availability of data and material*

The datasets used in the current study are available from the corresponding author on reasonable request.

## *Code availability*

The MLChem4D and QMLMaterial programs are available from the corresponding author on reasonable request.

## *Author contributions*

Conceptualization: MPL, HZH, MN, JH, DG, CS, DRS. Funding acquisition: MPL, DS, CS. Methodology: MPL, HZH, DS. Project administration: MPL, DRS. Resources: MPL, DRS. Software: MPL, HZH. Supervision: MPL, HZH, DRS, MN, DG, CS. Validation: MPL, HZH, JH, MN, DG. Writing – original draft: MPL. Writing – review & editing: MPL, HZH, JH, MN, DG, CS, DRS.

## *Acknowledgements*

The support of the Brazilian agencies: Fundação de Amparo à Pesquisa do Espírito Santo (FAPES), Conselho Nacional para o Desenvolvimento Científico e Tecnológico (CNPq) and Coordenação de Aperfeiçoamento de Pessoal de Ensino Superior (CAPES) are gratefully acknowledged. Work supported by the National Research Council of Canada (NRC), Artificial Intelligence for Design program and by the Natural Sciences and Engineering Research Council of Canada, Discovery Grant (RGPIN-2019-03976). We appreciate discussions with Sergey Gusarov, Khabat Heshami, Utkarsh Singh and Alain Tchagang from NRC.

# Supporting Information

## *Exploring Quantum Active Learning for Material Design and Discovery*

Maicon Pierre Lourenço[1*], Hadi Zadeh-Haghighi[2], Jiří Hostaš[3,4], Mosayeb Naseri[4], Daya Gaur[5], Christoph Simon[2] and Dennis R. Salahub[4]

1- Departamento de Química e Física – Centro de Ciências Exatas, Naturais e da Saúde – CCENS – Universidade Federal do Espírito Santo, 29500-000, Alegre, Espírito Santo, Brasil.
2- Department of Physics and Astronomy, Institute for Quantum Science and Technology, Quantum Alberta, and Hotchkiss Brain Institute, University of Calgary, Calgary, AB T2N 1N4, Canada.
3- Digital Technologies Research Centre, National Research Council of Canada, 1200 Montréal Road, Ottawa, ON, K1A 0R6 Canada.
4- Department of Chemistry, Department of Physics and Astronomy, CMS Centre for Molecular Simulation, IQST Institute for Quantum Science and Technology, Quantum Alberta, QHA Quantum Horizons Alberta, University of Calgary, 2500 University Drive NW, Calgary, AB, T2N 1N4, Canada.
5- Department of Computer Science, University of Lethbridge, 4401 University Dr. West Lethbridge, AB T1K 3M4, Canada.

*\* Address correspondence to:* [maiconpl01@gmail.com(MPL)](mailto:maiconpl01@gmail.com)*.*

*Statistical regression*

The idea behind statistical regression is to obtain N *observed* properties $\mathbf{y} = (y^{(1)}, ..., y^{(n)})$, i = 1, ...., N; to describe $\mathbf{y}$ statistically, the descriptor $\mathbf{x}^{(i)} = (x_1^{(i)}, ..., x_k^{(i)})$, with K variables is required. The property in our case is the piezoelectric coefficient and energy storage density for $Ba_{(1-x)}A_xTi_{(1-y)}B_yO_3$ perovskites and band-gap for $AA'BB'O_6$ perovskites. This results in a matrix $\mathbf{X}$ of dimension (N × K), called the feature matrix which is associated with the one-dimensional vector $\mathbf{y}$ (the objective function) of dimension (N). Here we define descriptor $\mathbf{x}^{(j)} = (x_1^{(j)}, ..., x_k^{(j)})$ in the virtual space ($N_{virtual}^k$): j = 1, ..., $N_{virtual}^k$; where *j* stands for the *j-th* virtual (non-observed) structure.

To model the desired problem in this manner, several surrogate models, such as support vector regression (SVR) or Gaussian Process regression (GP) can be used by utilizing high level libraries, such as scikit-learn[1].

After performing the regression, the statistical model is obtained and represented as:

$$\hat{y} = \hat{f}(X), \qquad (1)$$

where $\hat{y}$ is the vector with the predicted properties and $\hat{f}(X)$ is the statistical model (the predictor) designed from the SVR our GPR models.

Usually, to obtain a model without data bias, the matrices $\mathbf{X}$ and $\mathbf{y}$ – which define the initial data to obtain and test the ML models ($\mathbf{X}^l, \mathbf{y}^l$) – are split into two other matrices: ($\mathbf{X}^{train}, \mathbf{y}^{train}$), which are used to train the statistical model and ($\mathbf{X}^{test}, \mathbf{y}^{test}$) to validate it.

In order to obtain the average (μ($\mathbf{x}^{(i)}$) for the computed structures and μ($\mathbf{x}^{(j)}$) for the non-computed, virtual, structures) and the uncertainty (the standard deviation σ($\mathbf{x}^{(i)}$) and σ($\mathbf{x}^{(j)}$)), the matrices ($\mathbf{X}^{train}, \mathbf{y}^{train}$) are partitioned K times for K-fold cross-validation (CV) or B times for a non-parametric bootstrap (BS). For each partition p (in a total space of P=K or P=B) a statistical model is obtained. Hence: $\hat{y}^p = \hat{f}(X)$, p = 1, ..., P. Then, for each descriptor in the observed data set $\mathbf{x}^{(i)}$ (or for each descriptor j in the non-computed or virtual structures space: $\mathbf{x}^{(j)}$) the average $\mu(\mathbf{x}^{(i)})$ (or $\mu(\mathbf{x}^{(j)})$ for the virtual structures) and the standard deviation $\sigma(\mathbf{x}^{(i)})$ (or $\sigma(\mathbf{x}^{(j)})$) are obtained, as illustrated in figure S1 for the data in the observed space.

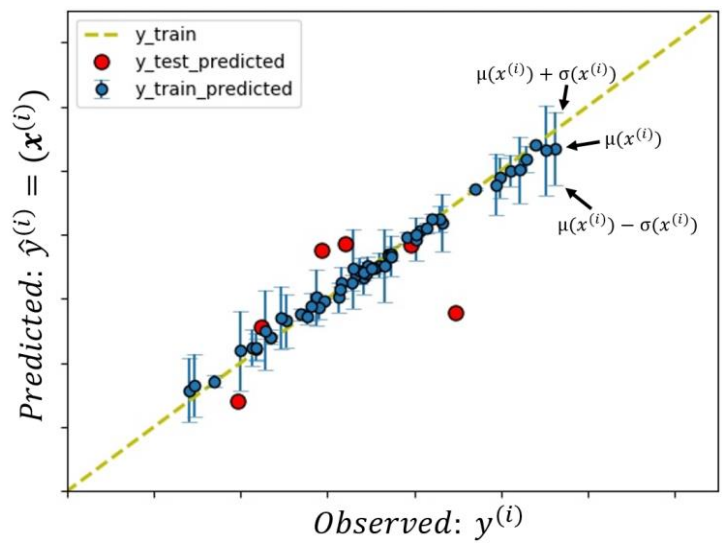

*Figure S1- Illustrative plot of the observed $y^{(i)}$ and predicted $\hat{y}(x^{(i)})$ target property. The use of SVR allows us to have a regression model for each data point $x^{(i)}$ represented from the mean $\mu(x^{(i)})$ and the standard deviation $\sigma(x^{(i)})$. The abscissa is the observed property and the ordinate the predicted one. The same for the unexplored space descriptors for inference and decision making: exploitation ($\mu(x^i)$) and exploration ($\sigma(x^{(j)})$).*

The mean $\mu(\mathbf{x}^{(j)})$ and the standard deviation $\sigma(\mathbf{x}^{(j)})$ for each descriptor entry in the non-observed (virtual) space (whose dimension is defined by $N_{virtual}^k$) will be used to obtain the acquisition function[2,3] which is used to indicate the next candidate to be computed. The next candidate is, then, incorporated in the initial descriptor matrix: ($\mathbf{X}^{train+1}$, $\mathbf{y}^{train+1}$) and the iteration process continues one step more until the optimization of the target property.

## *Quantum circuits used for data encoding*

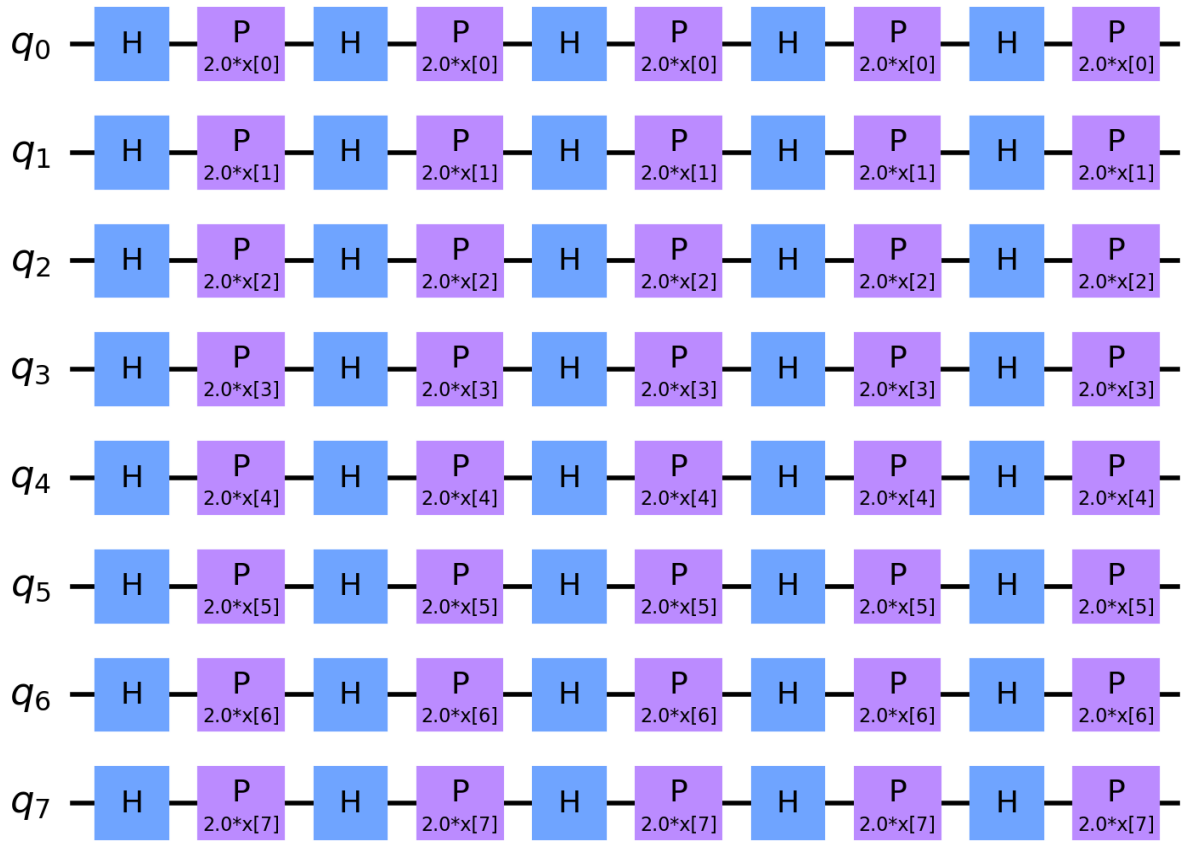

*Figure S2- ZFeatureMap quantum circuit (QC) encoding as implemented in Qiskit framework. This QC was built with 8 feature maps or qubits (q0, q1, ..., q7); the number of repeated circuits (reps) is equal to 5; the number of QC parameters, θ, is 8: x[0], ..., x[7].*

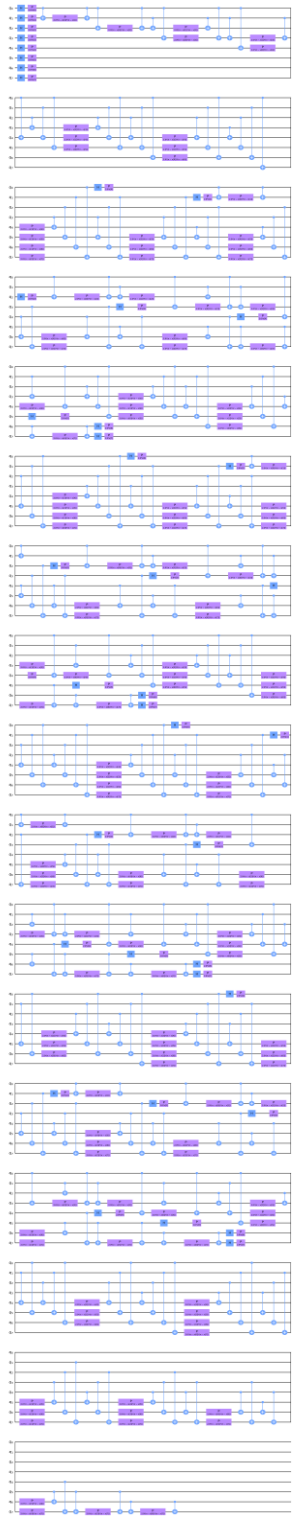

*Figure S3- ZZFeatureMap quantum circuit (QC) encoding with full entanglement, as implemented in Qiskit framework. This QC was built with 8 feature maps or qubits (q0, q1, ..., q7); the number of repeated circuits (reps) is equal to 5; the number of QC parameters, θ, is 8: x[0], ..., x[7].*

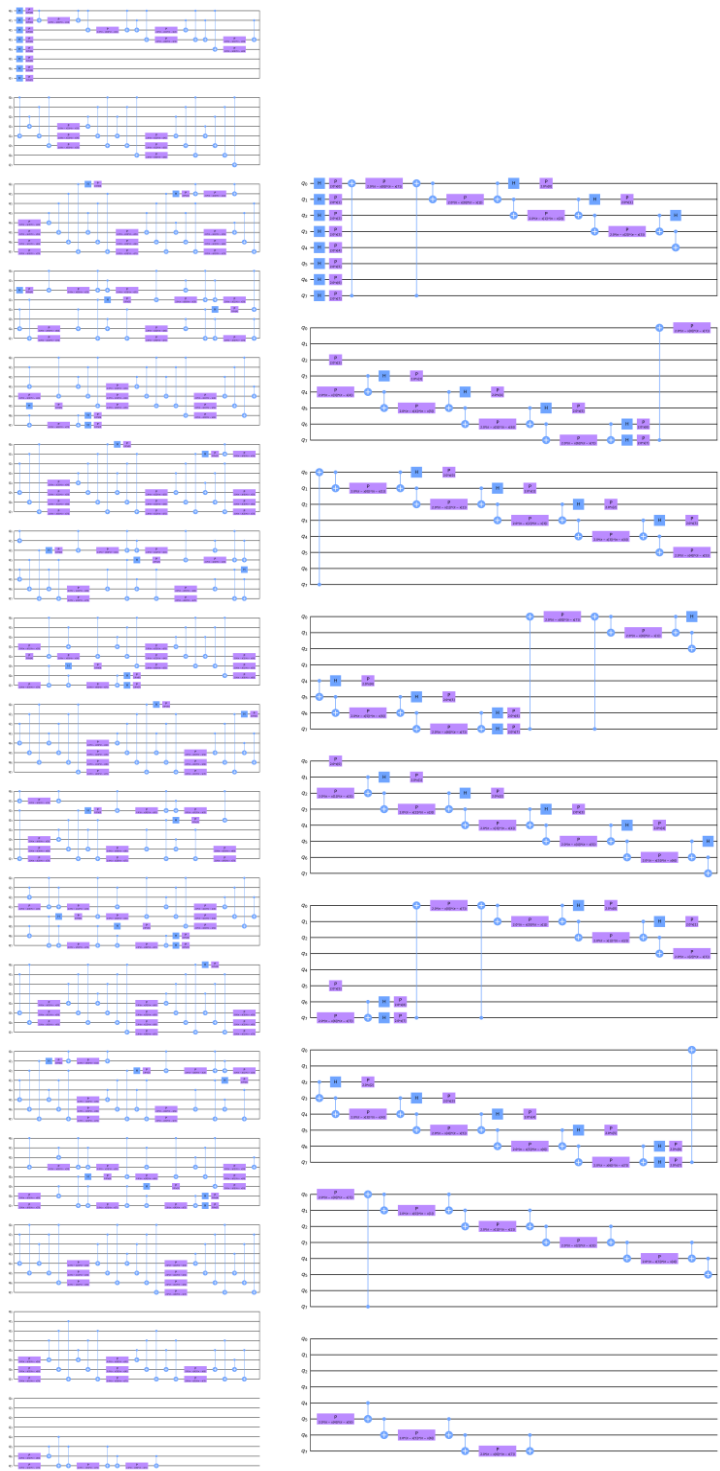

*Figure S4- PauliFeatureMap quantum circuit (QC) encoding with full entanglement on the left and circular entanglement on the right, as implemented in Qiskit framework. This QC was built with 8 feature maps or qubits (q0, q1, ..., q7); the number of repeated circuits (reps) is equal to 5; the number of QC parameters, θ, is 8: x[0], ..., x[7].*

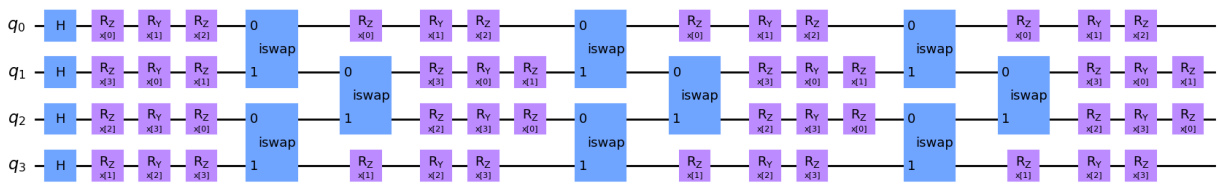

*Figure S5- HighDim quantum circuit (QC) encoding as defined in sQUlearn framework. This QC was built with 4 feature maps or qubits (q0, q1, ..., q3); the number of repeated circuits (reps) is equal to 4.*

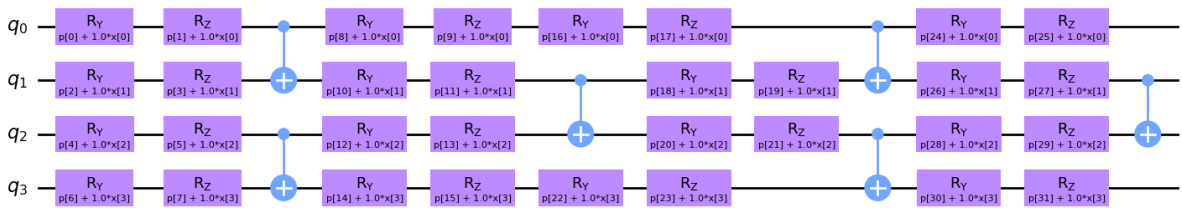

*Figure S6- YZ_CX quantum circuit (QC) encoding as defined in sQUlearn framework. This QC was built with 4 feature maps or qubits (q0, q1, ..., q3); the number of repeated circuits (reps) is equal to 4.*

## Grid search hyperparameterization for classical (SVR) and quantum (QSVR) machine learning models

The hyperparameters for SVR (using radial basis function kernel) and QSVR (using ZFeatureMap encoding and fidelity quantum kernel) were obtained using a gridsearch. After that, the best hyperparametes were selected and used in AL and QAL subsequent studies. This grid search is in line with the supporting information of Ref[4].

<u>*Piezoelectric coefficient of $Ba_{(1-x)}A_xTi_{(1-y)}B_yO_3$ perovskites (system I)*</u>:

Data size: 22.

C and gamma hyperparameters grid for classical SVR:

'svr__C': [1.0, 10, 20, 30, 40, 50, 60, 70, 80, 90, 100, 110, 120, 130, 140, 150, 200, **1000**]
'svr__gamma': [0.001, 0.007, 0.008, 0.009, 0.01, 0.02, 0.03, 0.04, 0.05, 0.06, 0.07, 0.08,

0.09, ***0.1***, 1, 10, 100]

C hyperparameter grid for QSVR using ZFeatureMap and fidelity quantum kernel:

**'qsvr__C':** [100, 200, 300, 400, 500, 600, 700, 800, 900, ***1000***, 1500, 2000]

The optimum hyperparameters are in bold.

*Band-gap of AA'BB'O$_6$ double perovskites (system II):*

Data size: 15.

C and gamma hyperparameters grid for classical SVR:

'svr__C': [***1.0***, 10, 20, 30, 40, 50, 60, 70, 80, 90, 100, 110, 120, 130, 140, 150, 200, 1000]
'svr__gamma': [***0.001***, 0.007, 0.008, 0.009, 0.01, 0.02, 0.03, 0.04, 0.05, 0.06, 0.07, 0.08, 0.09, 0.1, 1, 10, 100]

C hyperparameter grid for QSVR using ZFeatureMap and fidelity quantum kernel:

**'qsvr__C':** [100, 200, 300, 400, 500, 600, 700, 800, 900, ***1000***]

The optimum hyperparameters are in bold.

*Energy storage density of $Ba_{(1-x)}A_xTi_{(1-y)}B_yO_3$ perovskites (system III):*

Data size: 73.

C and gamma hyperparameters grid for classical SVR:

'svr__C': [0.1, 1.0, 10, 20, 30, 40, 50, 60, 70, 80, 90, 100, 110, 120, 130, ***140***, 150, 200, 1000]
'svr__gamma': [10.0**(-4), 0.001, 0.007, 0.008, 0.009, 0.01, 0.02, 0.03, 0.04, 0.05, ***0.06***, 0.07, 0.08, 0.09, 0.1, 1, 10, 100]

C hyperparameter grid for QSVR using ZFeatureMap and fidelity quantum kernel:

'qsvr__C': [100, 200, 300, 400, 500, 600, 700, 800, 900, ***1000***]

The optimum hyperparameters are in bold.

## *Grid search hyperparameterization for classical (GPR) and quantum (QGPR) machine learning models*

The hyperparameters for GPR (using "DotProduct + white" kernel) and QGPR (using HighDim encoding and projected quantum kernel) were obtained using a gridsearch. MBTR and spin multiplicity descriptors were used. Dimensionality reduction was then performed using PCA and the first 4 principal components were used as input features (descriptors) for both the GPR and the QGPR algorithms. Subsequently, the optimal hyperparameters were selected and used in the AL and the QAL studies. This grid search is in line with the supporting information of Ref[4].

*Structural and spin search determination of 3Al@Si$_{11}$ [2, 4, 6]:*

Data size: 60.

alpha hyperparameters grid for classical GPR:

'gpr_alpha': [1e15, 1e10, 1e5, 1e3, **10,** 1, 1e-1, 1e-3, 1e-5, 1e-10, 1e-15]

sigma hyperparameters grid for quantum GPRR:

'qgpr_sigma': [1e-6, 1e-5, 1e-4, **1e-3**, 1e-2, 1e-1, 1e0, 1e1, 1e2, 1e3]

The optimum hyperparameters are in bold.

## *Quantum machine learning set up*

The regression mean ($\mu$) and its epistemic uncertainty ($\sigma$) used in QAL were obtained from QSVR models, where the $\mu$ and $\sigma$ were obtained by CV with 5-fold resampling (K = 5): CV5. From those, the acquisition function EI (Eqs. 8 and 9 in the main text) was evaluated to indicate the next perovskites to have their properties investigated or measured. The QSVR supervised learning regression algorithm (using different QC encoding to obtain the fidelity quantum kernel, FQK) is available in the current version of MLChem4D, which was developed in Python3.x and uses the scikit-learn library as well as Qiskit for quantum computing and QML. In the design loop, the ML regression fit is made on 95 % of the observed data (the training set) and tested on 5 % of it (the test set).

Tab. 1 presents the mean absolute error for the training and testing set of the systems (I), (II) and (III) obtained by different AL and QAL iterations. 5-fold cross validation was used. Also, the optimum hyperparameters are shown. They were obtained by grid search for a fixed perovskite database, as described in detail here in the SI.

*Table 1. Statistics of the hyperparameters used in this work, as defined in the scikit-learn and Qiskit libraries using the ZFeatureMap for data encoding for the modified perovskites. 95 % of the data were randomly chosen for the training set and 5 % for testing for all materials studied, as defined during the QAL cycles. The cost function used to evaluate the quality of the regression was the mean absolute error (MAE) and it is presented for different data sizes obtained by QAL. "Hyper." means: hyperparameters used in the QSVR (using fidelity quantum kernel, FQK) or classical SVR (radial basis function, RBF, kernel) regression. The MAE for $Ba_{(1-x)}A_xTi_{(1-y)}B_yO_3$ piezoelectric coefficient (system I), band-gap of $AA'BB'O_3$ double-perovskites (system II) and energy storage density of $Ba_{(1-x)}A_xTi_{(1-y)}B_yO_3$ (system III) has the following units: pC/N, eV and mJ/cm$^3$, respectively. Quantum circuit features: "S" is superposition; "E" is entanglement; "NQ" is non-quantum (classic).*

| System I | Hyper. | Quantum features | MAE train | MAE test | MAE train | MAE test | MAE train | MAE test |
|---|---|---|---|---|---|---|---|---|
| | | | 22 data | | ~32 data | | ~42 data | |
| $Ba_{(1-x)}A_xTi_{(1-y)}B_yO_3$ | C = 10$^3$; ZFeatureMap; kernel = FQK | S | 14.06 | 90.04 | 10.91 | 73.14 | 25.81 | 79.35 |
| | C = 10$^3$; ZZFeatureMap; kernel = FQK | S + E | 12.82 | 82.05 | 26.03 | 88.81 | 26.13 | 59.01 |

| | Hyper. | | MAE train | MAE test | MAE train | MAE test | MAE train | MAE test |
|---|---|---|---|---|---|---|---|---|
| | C = 10³; PauliFeatureMap; kernel = FQK | S + E | 15.82 | 13.62 | 20.51 | 89.08 | 22.90 | 68.82 |
| | C = 10³; γ = 10⁻¹; kernel = RBF | NQ | 16.25 | 87.11 | 25.65 | 59.16 | 24.84 | 80.91 |
| **System II** | **Hyper.** | | **MAE train** | **MAE test** | **MAE train** | **MAE test** | **MAE train** | **MAE test** |
| | | | **10 data** | | **~20 data** | | **~30 data** | |
| **AA'BB'O₆** | C = 10³; ZFeatureMap; kernel = FQK | S | 0.15 | 0.75 | 0.18 | 0.42 | 0.16 | 0.87 |
| | C = 10³; ZZFeatureMap; kernel = FQK | S + E | 0.16 | 0.70 | 0.16 | 0.67 | 0.14 | 0.47 |
| | C = 10³; PauliFeatureMap; kernel = FQK | S + E | 0.08 | 0.84 | 0.16 | 0.21 | 0.16 | 0.54 |
| | C = 1; γ = 10⁻³; kernel = RBF | NQ | 0.16 | 0.11 | 0.22 | 0.49 | 0.21 | 0.36 |
| **System III** | **Hyper.** | | **MAE train** | **MAE test** | **MAE train** | **MAE test** | **MAE train** | **MAE test** |
| | | | **73 data** | | **~93 data** | | **~113 data** | |
| | C = 10³; ZFeatureMap; kernel = FQK | S | 2.06 | 6.31 | 1.93 | 4.68 | 1.93 | 9.95 |
| **Ba₍₁₋ₓ₎AₓTi₍₁₋** | C = 10³; PauliFeatureMap; kernel = | S + E | 1.80 | 16.08 | 2.05 | 13.46 | 2.06 | 13.92 |
| y)$B_yO_3$ | FQK | | | | | | | |
| | C = 140; γ = 0.06; kernel = RBF | NQ | 4.77 | 7.30 | 4.40 | 5.32 | 4.43 | 5.86 |

The regression QSVR and SVR hyperparameters that resulted in the smallest mean absolute error (MAE) for the training and the testing set are shown in Tab. 1. The regularization parameter $C$ for QSVR was obtained considering the ZFeatureMap data encoding for certain perovskites data (fixed) and was used in further QAL studies with ZZFeatureMap and PauliFeatureMap data encoding with entanglement – as shown in Tab. 1. For classical SVR (used in AL), the optimum hyperparameters presented in Tab. 1 – regularization parameter ($C$) and the kernel coefficient ($\gamma$) for the radial basis function kernel (RBF) – were optimized in a grid search (also for a fixed data) with tolerance factor equal to $10^{-3}$ and ε = 0.01. The optimum hyperparameters found were used subsequently in the AL (classical) study, as shown in Tab. 1. Further information can be found in the documentation of the scikit-learn library. Also, additional information about statistical regression, classical (RBF) and quantum (FQK) kernels can be found here in the SI.

*Table 2. Statistics of the hyperparameters used in this work, as defined in the scikit-learn and sQUlearn libraries for the 3Al@Si$_{11}$ nanoparticle. 95 % of the data were randomly chosen for the training set and 5 % for testing for 3Al@Si$_{11}$, as defined during the QAL cycles using exploitation for decision making. The cost function used to evaluate the quality of the regression was the mean absolute error (MAE) and it is presented for different data sizes obtained by QAL. "Hyper." means: hyperparameters used in the QGPR (using projected quantum kernel, PQK, or fidelity quantum kernel, FQK) or classical GPR (radial basis function, RBF, kernel) regression. The MAE is in Hatree × $10^{-2}$. The MBTR descriptor was used and PCA was applied, allowing a dimensionality reduction of 4. Quantum circuit features: "S" is superposition; "E" is entanglement; "NQ" is non-quantum (classic).*

| System | Hyper. | Quantum features | MAE train | MAE test | MAE train | MAE test | MAE train | MAE test |
|---|---|---|---|---|---|---|---|---|

| | IV | | | | | | | |
|---|---|---|---|---|---|---|---|---|
| | | | | **20 data** | | **~100 data** | | **~200 data** | |
| **3Al@Si$_{11}$** | featureMap = HighDim; Kernel = PQK; σ=0.001 | S + E | 0.12 | 7.90 | 0.28 | 3.36 | 0.43 | 3.48 |
| | featureMap = YZ_CX; Kernel = PQK; σ=0.001 | S + E | 0.12 | 6.58 | 0.31 | 2.74 | 0.42 | 3.10 |
| | Kernel = DotProduct + WhiteKernel, σ[1.0, ($10^{-3}$-$10^3$)], noise[10.0, ($10^{-3}$-$10^3$)]; α = 10 | NQ | 1.15 | 0.07 | 2.96 | 3.62 | 3.06 | 4.00 |

The GPR and QGPR hyperparameters used during the QAL – as defined in scikit-learn and sQUlearn) that resulted in the smallest mean absolute error (MAE) for the training and the testing set are shown in Tab. 2.

## *Single-perovskite descriptor*

The set of descriptors for non-stoichiometric perovskites – as for $Ba_{(1-x)}A_xTi_{(1-y)}B_yO_3$ for the energy storage density search – is composed of the weighted atomic properties ($\bar{P}_J$): $\bar{P}_J = \sum_J^{N_{ions}} x_J P_J$; where $x_J$ is the fraction of ions that compose the sites A or B of the ABO$_3$ perovskite and $P_J$ is the property of the Jth ion at the A or B site. $N_{ions}$ is the number of ions in the A or B sites. Also, the following tolerance factors[5] are used as descriptor: $t_f = (\bar{r}_A + r_O)/[\sqrt{2}(\bar{r}_B + r_O)]$ and $t_t^{new} = \left(\frac{\bar{r}_O}{\bar{r}_B}\right) - Q_A \left[\left(Q_A - \frac{\bar{r}_A}{\bar{r}_B}\right) / \ln\left(\frac{\bar{r}_A}{\bar{r}_B}\right)\right]$ ; where $Q_A$ is the oxidation state of site A and $\bar{r}_A$, and $\bar{r}_B$ are the average atomic radius of the A and B sites, respectively, obtained by $\bar{P}_J$; $r_O$, is the ionic radius of oxygen.

The set of descriptors developed in this work is composed of a list with the following elements:

(I) The tolerance factors $t_f$ and $t_t^{new}$.

(II) The following properties of the A ($\bar{P}_A$) and B ($\bar{P}_A$) sites of ABO$_3$: (1) Shannon ionic radius[6,7]; (2) ideal bond length of A-O and B-O; (3) electronegativity; (4) van der Waals radius; (5) first ionization energy; (6) molar volume; (7) atomic number; (8) atomic mass. As they are for both the A and B sites, 2×8 = 16 features result.

(III) The properties related to the A and B sites, as described in II, divided ($\bar{P}_A/\bar{P}_B$) and multiplied ($\bar{P}_A . \bar{P}_B$); resulting in another 16 features. Therefore, the set of descriptors developed is composed of a list with 34 properties (2 + 16 + 16).

The aforementioned atomic properties used to create sets of descriptors for the perovskites were obtained from the Python Materials Genomics[8] (pymatgen) library.

## *Double-perovskite descriptor*

The descriptor for double-perovskites compounds is based on weighted atomic properties described in the previous section: SI-4. Since we need to include the information about both: the parent ABO$_3$ elements and the doped A' and B' sites, it results in effectively doubling the size of the descriptor when compared to the single perovskites. More specifically, 8 properties for each A, A', B, and B' element lead to a feature vector of length 32. Next, the properties were divided ( $\bar{P}_A/\bar{P}_B, \bar{P}_{A'}/\bar{P}_{B'}$ ) and multiplied ($\bar{P}_A . \bar{P}_B, \bar{P}_{A'} . \bar{P}_{B'}$) effectively doubling the size (64) of the descriptor. More information about this descriptor can be found in our ICLR 2023 Workshop publication[9] and the dataset can be downloaded at its associated Github page (https://github.com/jiri-hostas/EDA-and-ML-for-Perovskites). The materials follow the $A_xA'_{(1-x)}B_yB'_{(1-y)}O_6$ formula, where:

A = [Ba, Bi, Ca, Dy, Eu, Gd, La, Li, Lu, Na, Pr, Sm, Sr, Tb, Y, Cu, Mg]

and

B = [Ag, Al, Bi, Ce, Co, Cr, Dy, Eu, Fe, Ga, Gd, In, La, Mg, Mn, Nd, Ni, Pr, Sc, Sm, Ta, Ti, Zn, I, Mo, Nb, Te, V, W]

### *Spin multiplicity and structural descriptors*

The spin-multiplicity descriptor (SMD) is represented by the vector $[(2S + 1), S, 2S \times (S + 1)^{1/2}, n_{e\text{-unpaired}}]$, where S is the total spin quantum number; $(2S + 1)$ is the multiplicity; $2S \times (S + 1)^{1/2}$ is the spin magnetic moment; $n_{e\text{-unpaired}}$ is the number of unpaired electrons. Take, for example, a cluster with SM equal to 4 (a quartet), we have: $(2S + 1) = 4$; $S = 3/2$; $2S \times (S + 1)^{1/2} = 3.87$; $n_{e\text{-unpaired}} = 3$. Thus, the vector is: [4, 1.5, 3.87, 3]. The SMD descriptor is simply added to the structural-feature vector so that it can be used in supervised ML or QML where both structural and electronic structure descriptors are incorporated. Then, the resulting vector is transformed by principal component analysis (PCA), as implemented in scikit-learn. Thus, the spin multiplicities (SMs) are accounted for inside the AL or QAL loop for the structural determination.

The many-body tensor representation descriptor (MBTR) was used to represent the structures of the $3Al@Si_{11}$ nanoparticle isomers or homotops. The MBTR is implemented in the DScribe[10] library which is interfaced in QMLMaterial[11]. MBTR descriptor encodes information about the whole atomic structure and can be used to predict properties involving molecular energies. In the MBTR, a configuration of k atoms is transformed into a single scalar by using a geometry function $g_k$. Then kernel density estimations with a Gaussian kernel are employed to expand the scalar value.

## Tutorial video explaining how to use the MLChem4D and QMLMaterial programs

YouTube video explaining how to run the classical Active Learning (AL) and Quantum Active Learning (QAL) for piezoelectric coefficient maximization of $ABO_3$ modified perovskites using SVR and QSVR: https://youtu.be/jLhFajQ0Uus. Example for running QMLMaterial with classical AL and QAL for $3Al@Si_{11}$ with spin multiplicities 2, 4 and 6 using GPR and QGPR: https://youtu.be/lVFI_XZxcyo.